# A mathematical theory of fame

M.V. Simkin and V.P. Roychowdhury

*Department of Electrical Engineering, University of California, Los Angeles, CA 90095-1594*

**Abstract.**  We study empirically how the fame of WWI fighter-pilot aces, measured in numbers of web pages mentioning them, is related to their achievement, measured in numbers of opponent aircraft destroyed. We find that on the average fame grows exponentially with achievement; the correlation coefficient between achievement and the logarithm of fame is 0.72.  The number of people with a particular level of achievement decreases exponentially with the level, leading to a power-law distribution of fame. We propose a stochastic model that can explain the exponential growth of fame with achievement.  Next, we hypothesize that the same functional relation between achievement and fame that we found for the aces holds for other professions. This allows us to estimate achievement for professions where an unquestionable and universally accepted measure of achievement does not exist. We apply the method to Nobel Prize winners in Physics. For example, we obtain that Paul Dirac, who is a hundred times less famous than Einstein contributed to physics only two times less. We compare our results with Landau's ranking.

## How fame depends on achievement

For almost all professions it is hard to define an objective measure of achievement.  As a result, the question of its relationship to fame is ill posed.   Fortunately, there is at least one case where an unquestionable measure of achievement does exist. This is the case of fighter-pilots, for whom the achievement is measured as a number of opponent aircraft destroyed. A fighter-pilot who achieved five or more victories is called an ace. The website [1] contains the names of all WWI aces together with the number of victories each of them had achieved.

In the Internet age there is an easily assessable index to fame:  the number of web pages (as found using Google) that mention the person in question [2].  We will refer to it as the number of *Google hits*.

In [3] we compared achievement and fame of 392 German WWI aces. The result is in Figure 1. The correlation coefficient between achievement and the logarithm of fame is 0.72. In contrast the correlation between achievement and fame (without logarithm) is only 0.48. The significance of the correlation coefficient, $r$, is that $r^2$ is the fraction of variance in the data which is accounted for by linear regression. This means that about half ( $0.72^2 \cong 52\%$ ) of the difference in fame is determined by the difference in achievement. Figure 2 shows the distributions of achievement and fame. One can see that achievement is distributed exponentially and that fame has a power-law tail.

We can show that when the distribution of achievement, $A$, is exponential,

$$p(A) = \alpha \exp(-\alpha \times A), \tag{1}$$

and fame, $F$, grows exponentially with achievement,

$$F(A) = C \exp(\beta \times A), \tag{2}$$

the fame is distributed according to a power law. Clearly, elimination of $A$ from Eq. (1) using Eq. (2) leads to:

$$p(F) = \frac{\alpha}{\beta} C^{\alpha/\beta} F^{-\gamma}; \gamma = 1 + \frac{\alpha}{\beta}. \tag{3}$$

This derivation follows the one, used by Yule to explain the power law in the frequency distribution of sizes of biological genera: see chapter 2 of Ref.4. After substituting the values $\alpha \cong 0.083$ and $\beta \cong 0.074$, obtained from the least-square fits of the data (see Figs. 1 and 2(a) into Eq.(3) we get $\gamma \cong 2.1$ which is quite close to $\gamma \cong 1.9$ obtained by fitting the actual distribution of fame (see Fig.2 (b) ).

Exponential growth of fame with achievement leads to its unfair distribution. With 80 confirmed victories, Manfred von Richthofen is the top-scoring ace of the WWI. With 4,720 Google hits[1] he is also the most famous. The total number of opponent aircraft destroyed by German aces in WWI is 5050. At the same time there are 17,674 Google hits for all of the German aces. This means that Manfred von Richthofen accumulated 27% of fame, while being personally responsible for shooting down only 1.6% of opponent aircraft. On the opposite side, 60 lowest scoring aces (with 5 victories each) together shot down 300 aircraft, or 5.9% of all aircraft destroyed. However, together they got only 463 Google hits, or 2.6% of fame.

On the other side, one may still be glad that there is a strong positive correlation between the logarithm of aces' fame and their achievement. The correlation coefficient between the scores received by different wines in blind tasting and the logarithm of their price is negative [5].

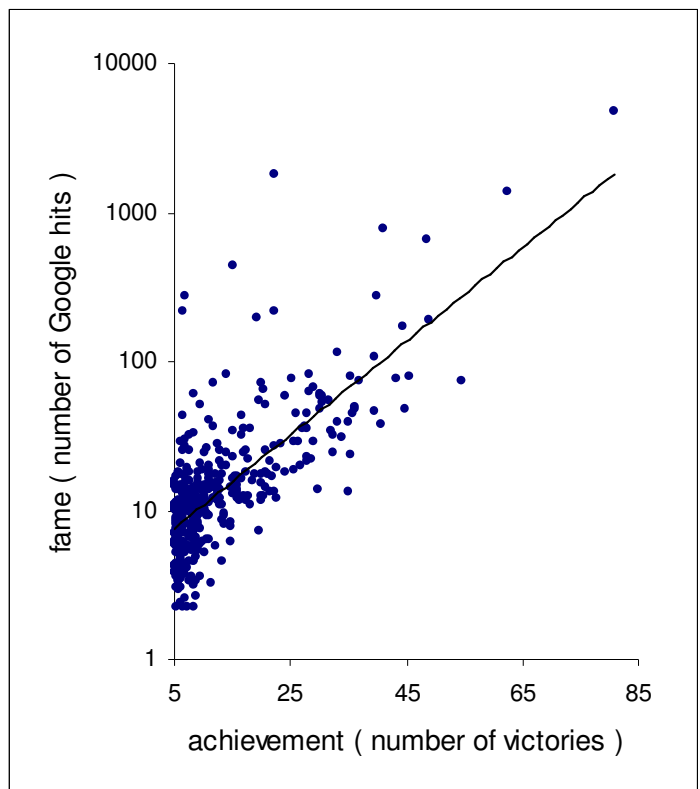

**Figure 1.** A scatter plot of fame versus achievement for 392 German WWI aces. The correlation coefficient of 0.72 suggests that $0.72^2 \cong 52\%$ of the variation in fame is explained by the variation in achievement. The straight line is the fit using Eq.2 with $\beta \cong 0.074$. There are many aces with identical values of both achievement and fame. To keep the density of dots representing the density of data points we added random numbers between zero and one to every value of achievement and fame.

---

[1] We collected the data used during the work on Ref. [3] that is in 2003. Today's numbers of Google hits are different.

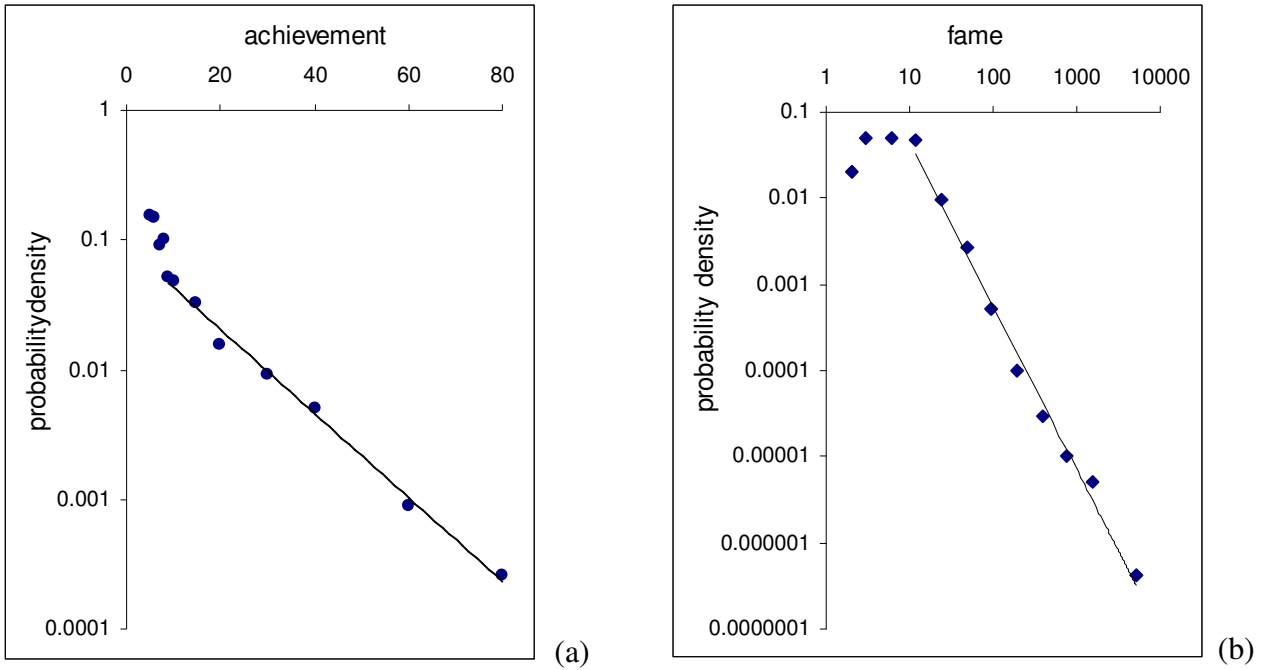

**Figure 2.** **(a)** The distribution of achievement (number of victories) for 392 German WWI aces. The straight line is the fit using Eq.1 with $\alpha \cong 0.083$. **(b)** The distribution of fame (number of Google hits) for the same aces. The straight line is the fit $p(F) \propto F^{-\gamma}$ with $\gamma \cong 1.9$.

## Stochastic modeling of fame

A simple stochastic model can explain why fame grows exponentially with achievement. It is convenient to describe the dynamics of fame in terms of memes [6] (we use this word in the sense of a piece of information, which can pass from one mind to another). We define the fame of X as the number of people who know X, or, in other words, the number of memes about X. In practice we can't count the number of memes, but we can count the number of webpages. It is natural to assume that the number of webpages, mentioning X, is proportional to the number of memes.

The rate of encountering memes about X is obviously proportional to the current number of such memes in the meme pool. We will assume that when someone meets a meme about X, the probability that it will replicate into his mind is proportional to X's achievement (which thus plays the role of meme's Darwinian fitness). The rate of the spread of a meme about someone with achievement $A$ is thus:

$$s = \nu A \ . \tag{4}$$

Here $\nu$ is an unknown independent of $A$ coefficient, which comprises the effects of all factors other than achievement on meme spread. The expectation value of the number of memes obeys the following evolution equation:

$$\frac{d\langle F \rangle}{dt} = s\langle F \rangle = \nu A\langle F \rangle \ . \tag{5}$$

If at time 0 there was only one copy of the meme the solution of Eq. (5) is

$$\langle F(t) \rangle = \exp(\nu t \times A), \tag{6}$$

which is Eq.(2) with

$$\beta = \nu t \,. \tag{7}$$

In the case of the aces, $t$ is time passed since WWI. After substituting Eq.(7) into Eq.(3), we get:

$$\gamma = 1 + \frac{\alpha}{\nu t}$$

The value of $\gamma$ is consistent with the experimental data on aces if $\nu t \approx \alpha$.

## Estimating achievement from fame

Given the value of fame, we can estimate achievement by inverting Eq. (1) [7]:

$$A(F) = \ln(F/C)/\beta \tag{8}$$

We computed for every ace an estimate of achievement based on his fame using Eq.(8). We then divided it by his real achievement. Figure 3 shows the distribution of such ratios of estimated and real achievements for all 392 German WWI aces. With high accuracy, we can approximate it by a lognormal distribution with mean zero and variance of 0.49. Kolmogorov-Smirnov test is passed with the p-value of 0.40. Analysis of the data of Fig.3 shows that with 50% probability estimated achievement is between 0.7 and 1.44 of the real achievement. With 95% probability, estimated achievement is between 0.43 and 2.4 of real achievement. And with the 85% probability the real achievement is between two times more and two times less than the estimate. The estimate is thus not very accurate; however, even such crude estimate gives an insight in the fields where we have no clue of how to measure achievement.

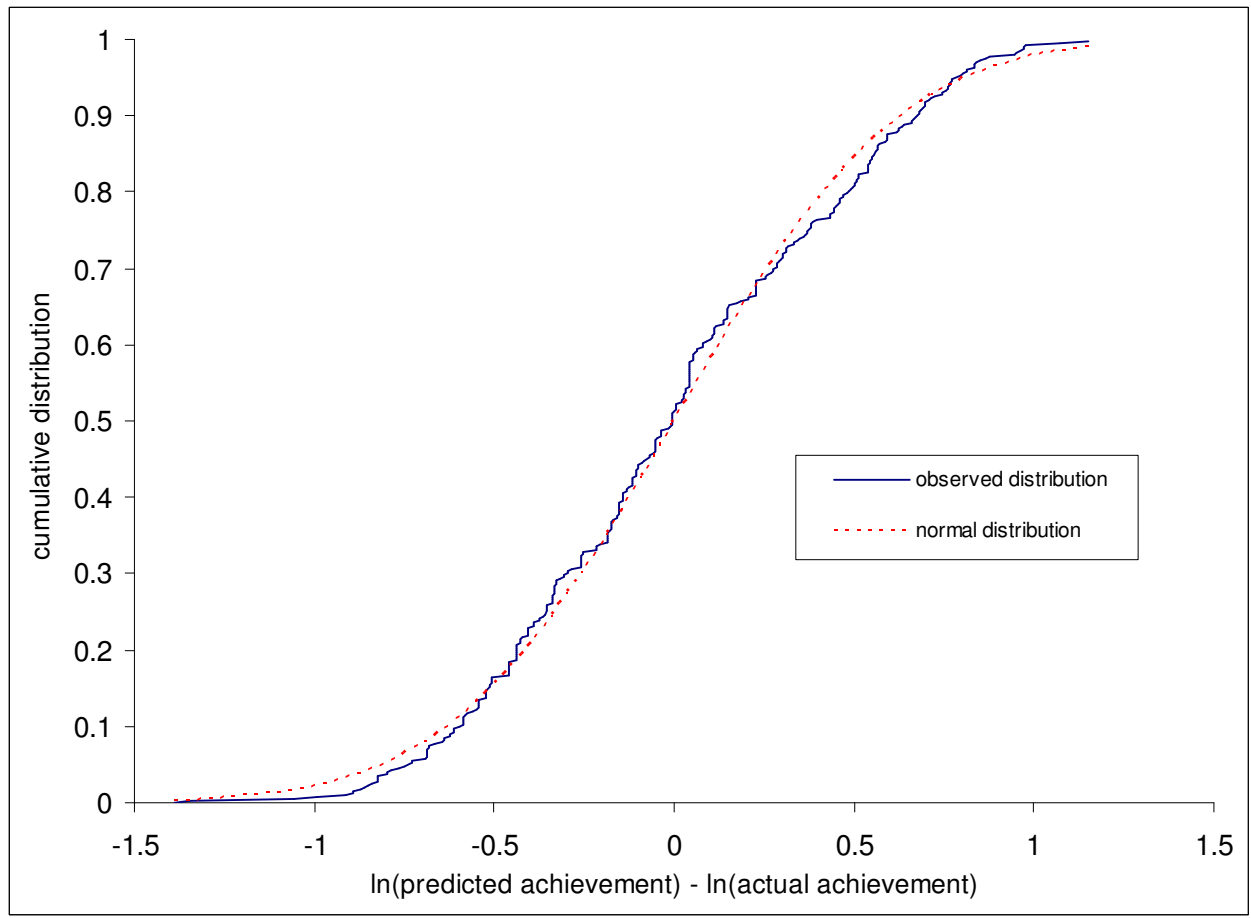

**Figure 3.** Cumulative distribution of the ratio of estimated achievement to actual achievement.

After the publication of our original paper [3] on the relationship between achievement and fame of fighter-pilot aces, Bagrow *et al* reported [8], that for physicists the relation between achievement and fame is linear. However, they used the number of published papers as a measure of achievement. Recent Bogdanoff affair [9] had shown that one could publish in respectable journals even papers consisting of an incoherent stream of buzzwords of modern physics. Thus, we cannot use the number of published papers to measure scientific achievement. Garfield suggested [10] that the number of citations to scientist's papers is the true measure of scientific achievement. In another study [11], [12] we had shown that since citations multiply by mere copying this measure is also questionable. While the number of citations may be increasing with the size of scientific contribution made in the paper, it is not obvious what the exact relation between these variables is. Here we hypothesize that the same exponential relation between fame and achievement, as we found for fighter pilots, holds for people of other professions. We then use their fame (measured in Google hits) to infer their achievement. Note that we do not say that web hit counts are preferable to citation counts. These two measures of fame are strongly correlated and are interchangeable. We used web hits because we used them for fighter pilots aces. The point is not that one should use web hits, but that one should take a logarithm of fame to estimate achievement.

Let us now try to estimate physicist's achievement based on their fame. Table 1 shows the names of 45 pre-WWII Nobel Laureates in Physics[2], ranked according to their fame. Figure 4 shows their fame distribution. It is very similar to the fame distribution of aces (see Fig. 2(a)). We hypothesize that the relation between achievement and fame for physicists is, similar to aces, given by Eq. (8). A big difference with the case of aces is that we do not know the values of $\beta$ and $C$. For the case of aces, where we knew achievement values, we determined these coefficients by regression. For physicists since we do not know the achievement (we actually are to determine it) the coefficients are unknown. The fact that $\beta$ is unknown is irrelevant, as it cancels out from the ratio of achievements.

---

[2] The list includes all of the pre-WWII Nobel Laureates in Physics, excluding Charles Wilson who had so many namesakes that his fame was impossible to determine.

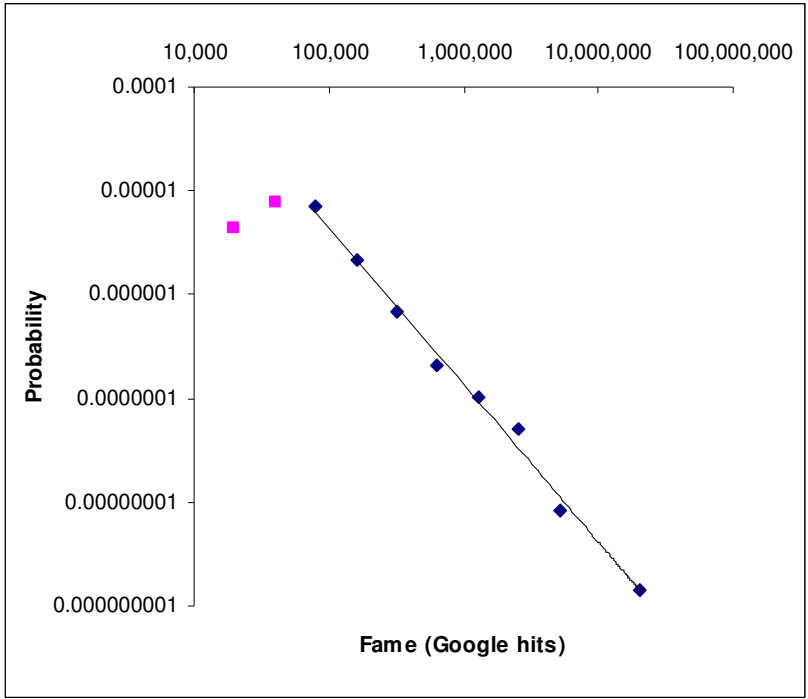

**Figure 4.** Distribution of fame of Nobel Prize winning physicists. The solid line is a power-law fit with exponent 1.5. This distribution is very similar to the fame distribution of flying aces given Figure 2(b).

The most famous physicist in Table 1 is Albert Einstein, according to Eq.(8), he is, most likely, the most achieved. Therefore, we will use him as a unit of achievement, which we denote as $A_E$. From Eq.(8) we then get:

$$\frac{A}{A_E} = \frac{\ln(F/C)}{\ln(F_E/C)} \qquad (9)$$

We still need to know $C$ to find the achievement in Einsteins. While exact determination of $C$ is impossible, we can find an upper bound on it. It is the fame of the least famous person in the list: $C$ cannot be more than that because in that case the achievement of the least famous person will become negative. The least famous person on our list is Nils Dalén. His Nobel Prize is also the most contested: many believe his achievement is not worthy of it. Dalén received Nobel Prize for his invention of the automatic sun valve, which regulates a gaslight source by the action of sunlight, turning it off at dawn and on at dusk. At the same time, most of the things invented by other people from our list have no practical applications, and those, which have applications, are very dangerous. Nevertheless, we will side with the contestants and assign Dalén the achievement of 0. Then we can substitute Dalén's fame, $F_D$, for $C$:

$$\frac{A}{A_E} \approx \frac{\ln(F/F_D)}{\ln(F_E/F_D)} \qquad (10)$$

Eq.(10) is an estimate of the lower bound on the achievement in Einsteins. This is because $C \le F_D$ and when $C < F_D$ Eq.(9) will give a higher value for $\frac{A}{A_E}$ than Eq.(4) for everyone but Einstein.

The estimates of achievement, computed using Eq.(10) are given in Table 1. We should note that the data presented in the table is very noisy since some physicists got additional fame for reasons other then their scientific achievement, for example for their role in public life. However, similar things happened to fighter-pilot aces. For example, Max Immelmann got additional web hits for invented by him aerobatic maneuver called "Immelmann Turn." Hermann Göring got additional web hits for his political activity. He is the second German WWI ace fame wise, though with his 22 victories he is only on the 60th place according to his achievement. The data shown in Fig.3 include all such cases. Let us emphasize that the error boundaries of the estimate of achievement from fame are based on the data that include all the noise and the extra hits received by the aces for activities other than their career as a fighter pilot.

One objection that we encountered is that Max Planck got a lot of fame due to the singular event: renaming of Keiser Wilhelm Society into Max Planck Society. All the institutes under auspices of the society became Max Planck institutes. Every scientific paper published by the members of Max Planck institutes automatically mentions Max Planck in its address line. Similarly, when a news article or a blog entry discusses a discovery by a member of one of the institutes, it mentions scientist's affiliation and therefore Max Planck. Together they contribute a large share of web hits. A Google search for "Max Planck Institute" OR "Max Plank Institut" produces 6,500,000 hits. If we subtract this number from the total number of hits, we are left with 4,100,000. This shifts Max Planck from the second place to the third. The estimate of his achievement in Einsteins drops from 0.91 to 0.8 or by 12%. The effect is thus not very big.

Another critic said that using our method he estimated the achievement of one modern physicist, who is also a blogger, and obtained 0.65 Einsteins. However, this is exactly why we limited our study to pre-WWII Nobel Prize winners. If we included modern physicists then the dead would have to compete with the living that blog, write popular articles, and appear in TV-shows. Moreover, people not only prefer to talk about their living contemporaries to talking about people from the past. They also talk more about people from recent past than of people from distant past. Thus to make the comparison meaningful we need to compare people of the same epoch, which we did in our study.

The estimate of achievement of every physicist listed in Table 1 (with the only exception of Dalen) is at least 15% of Einstein's achievement. For example, Dirac and Schrödinger who are 90 and 60 times less famous than Einstein appear to achieve only two times less. This may seem shocking to some people. Are these results meaningful?

Half a century ago a Nobel Prize winning physicist Lev Landau classified theoretical physicists according to their achievement using a logarithmic scale [13], [14]. According to his ranking system, a member of the lower class achieved ten times less than a member of the preceding class. He placed Einstein in ½ class. In the $1^{st}$ class he placed Bohr, Schrödinger, Heisenberg, Dirac, Fermi, and de Broglie[3]. Thus, he thought that Einstein contributed to Physics $\sqrt{10} \approx 3$ times more than Dirac or Schrödinger. This is close enough to our estimate, according to which Einstein achieved 2 times more than Dirac or Schrödinger. Taking into account our errors of two times more or two times less, this agreement is perfect. Note that Landau's ranking is incomparably closer to our estimate than to a naïve estimate equating fame and achievement. The agreement becomes worse in the cases of Heisenberg and Bohr where we estimate that they achieved 0.6 and 0.7 Einsteins respectively. However, earlier in his life, during 1930s, Landau used another classification [13]. According to it Lorentz, Planck, Einstein, Bohr, Heisenberg,

---

[3] These are the only people from Table 1, whose Landau rankings were given in [13] or [14].

Schrödinger, Dirac all belonged to the $1^{st}$ class. Our results are compatible with this earlier Landau's classification.

Surowiecki argues [15] that if you take an average of the guesses of very many people you get an estimate, which is as good as or even better than an expert opinion. Every webpage about a particular person expresses its creator's opinion that the person in question is worthy of it. In our model, the number of people considering mentioning a particular person in their webpages is proportional to the current number of webpages mentioning the person in question. We assumed that the probability for the decision to be positive is proportional to the person's achievement. So the achievement obtained using our model is proportional to the fraction of positive judgments. Thus, the fact that our estimate of achievement of Nobel Prize winning physicist based on statistical analysis of the numbers of webpages mentioning them agrees fairly well with expert's (Landau's) opinion may be another demonstration of wisdom of crowds.

**Table 1**

| Physicist | Alternative names used in Google search, all joined using OR | June 2008 Google hits | Log over Dalen | Lower bound on the most likely achievement (in Einsteins) | Landau rank, where known |
|---|---|---|---|---|---|
| ALBERT EINSTEIN | | 22,700,000 | 8.53 | 1 | 0.5 |
| MAX PLANCK | MAX KARL ERNST LUDWIG PLANCK | 10,600,000 | 7.77 | 0.911 | |
| MARIE CURIE | | 6,300,000 | 7.25 | 0.850 | |
| NIELS BOHR | | 1,890,000 | 6.04 | 0.709 | 1 |
| ENRICO FERMI | | 1,730,000 | 5.95 | 0.698 | 1 |
| GUGLIELMO MARCONI | | 1,110,000 | 5.51 | 0.646 | |
| WERNER HEISENBERG | | 987,000 | 5.39 | 0.632 | 1 |
| ERWIN SCHRÖDINGER | ERWIN SCHROEDINGER | 375,000 | 4.43 | 0.519 | 1 |
| PIERRE CURIE | | 330,000 | 4.30 | 0.504 | |
| WILHELM RÖNTGEN | WILHELM CONRAD RÖNTGEN WILHELM CONRAD ROENTGEN WILHELM ROENTGEN | 272,000 | 4.10 | 0.481 | |
| PAUL DIRAC | PAUL ADRIEN MAURICE DIRAC PAUL AM DIRAC | 255,000 | 4.04 | 0.474 | 1 |
| LOUIS DE BROGLIE | LOUIS-VICTOR DE BROGLIE | 201,000 | 3.80 | 0.446 | 1 |
| LORD RAYLEIGH | LORD JOHN WILLIAM STRUTT RAYLEIGH | 167,000 | 3.62 | 0.424 | |
| MAX VON LAUE | | 142,000 | 3.45 | 0.405 | |
| HENDRIK LORENTZ | HENDRIK ANTOON LORENTZ | 119,000 | 3.28 | 0.384 | |
| ROBERT MILLIKAN | ROBERT ANDREWS MILLIKAN | 112,000 | 3.22 | 0.377 | |
| JAMES FRANCK | | 109,000 | 3.19 | 0.374 | |
| JAMES CHADWICK | | 99,100 | 3.09 | 0.363 | |
| CHARLES GUILLAUME | CHARLES EDOUARD GUILLAUME | 89,900 | 3.00 | 0.351 | |
| LAWRENCE | | 89,500 | 2.99 | 0.351 | |
| ALBERT MICHELSON | ALBERT ABRAHAM MICHELSON | 76,600 | 2.84 | 0.333 | |
| WILLIAM LAWRENCE BRAGG | | 74,500 | 2.81 | 0.329 | |
| JOSEPH JOHN THOMSON | | 73,700 | 2.80 | 0.328 | |
| ANTOINE BECQUEREL | ANTOINE HENRI BECQUEREL | 70,300 | 2.75 | 0.323 | |
| ARTHUR COMPTON | ARTHUR HOLLY COMPTON | 66,800 | 2.70 | 0.317 | |
| WILHELM WIEN | | 52,600 | 2.46 | 0.289 | |
| GABRIEL LIPPMANN | | 49,300 | 2.40 | 0.281 | |
| JOHANNES VAN DER WAALS | JOHANNES DIDERIK VAN DER WAALS | 48,800 | 2.39 | 0.280 | |
| PIETER ZEEMAN | | 47,200 | 2.35 | 0.276 | |
| WILLIAM HENRY BRAGG | | 46,800 | 2.34 | 0.275 | |
| JOHANNES STARK | | 45,900 | 2.32 | 0.273 | |
| MANNE SIEGBAHN | KARL MANNE GEORG SIEGBAHN | 45,000 | 2.30 | 0.270 | |
| PHILIPP LENARD | PHILIPP EDUARD ANTON LENARD | 40,000 | 2.19 | 0.256 | |
| CARL FERDINAND BRAUN | KARL FERDINAND BRAUN | 40,000 | 2.19 | 0.256 | |
| GUSTAV HERTZ | | 37,800 | 2.13 | 0.250 | |
| HEIKE KAMERLINGH-ONNES | | 35,100 | 2.06 | 0.241 | |
| SIR GEORGE THOMSON | GEORGE PAGET THOMSON | 29,900 | 1.90 | 0.222 | |
| CLINTON DAVISSON | CLINTON JOSEPH DAVISSON | 29,100 | 1.87 | 0.219 | |
| JEAN BAPTISTE PERRIN | | 28,600 | 1.85 | 0.217 | |
| CARL DAVID ANDERSON | | 26,400 | 1.77 | 0.208 | |
| OWEN RICHARDSON | WILLANS RICHARDSON | 24,900 | 1.71 | 0.201 | |
| CHARLES BARKLA | CHARLES GLOVER BARKLA | 24,500 | 1.70 | 0.199 | |
| CHANDRASEKHARA RAMAN | CHANDRASEKHARA VENKATA RAMAN | 22,100 | 1.59 | 0.187 | |
| VICTOR FRANZ HESS | | 17,200 | 1.34 | 0.157 | |
| NILS DALÉN | NILS GUSTAF DALÉN NILS GUSTAF DALEN | 4,490 | 0.00 | 0 | |